\documentclass[12pt]{article}
\usepackage[cp1251]{inputenc}
\usepackage[russian,english]{babel}
\usepackage{amsfonts,amssymb}
\usepackage{graphicx}
\usepackage{amsmath, amssymb, graphics, setspace}
\usepackage{hyperref}

\begin{document}

\title{The model of the black hole enclosed in dust. The flat space case.}
\maketitle
\begin{center}
P.~Jaluvkova$^{1,2}$ \quad E.~Kopteva$^2$ \quad Z.~Stuchlik$^2$\\
       $^1${\it Joint Institute for Nuclear Research, Dubna, Russia} \\
       $^2${\it Institute  of  Physics,  Faculty  of  Philosophy  and  Science,
Silesian  University  in  Opava, Czech Republic} \\
\end{center}

\begin{abstract}
The model is constructed to describe the Schwarzschild-like black hole enclosed in the dust cosmological background. It is an exact solution of Einstein equations for spherically symmetric dust distribution, and is a special case of Lemaitre-Tolman-Bondi solutions. The motion of the test particle in the model is investigated in comoving coordinate frame. Observable velocity of the particle is found from geodesic equations. It is shown that chosen reference system does not allow to solve the problem of 'all or nothing' behavior.
\end{abstract}

\section{Introduction}
\label{intro}
Since the pioneer work by McVittie \cite{17} there were numerous attempts to construct the model of the black hole enclosed in the space which is not empty (see e.g. \cite{reviews},\cite{Krasinski} for reviews, and some analysis in \cite{KorkinaKopteva}). This problem concerns a wide set of research directions, including the thermodynamics of black holes \cite{Giddings},\cite{FirouzjaeeMansouri},\cite{FirouzjaeeEllis},\cite{FirouzjaeeMansouri2}, the black hole horizon dynamics \cite{FirouzjaeeMansouri3}, the influence of cosmological expansion on the evolution of local objects \cite{MoradiFirouzjaeeMansouri},\cite{FaraoniJacques} etc. 
The problem is a point for active discussions and is still far from its final resolution. New exact solutions probably can help here. 

In this work we focus on investigation of the exact solution of Einstein equations for spherically symmetric dust distribution which is a special case of class of Lemaitre-Tolman-Bondi (LTB) solutions. It was recently obtained by use of mass function method \cite{KorkinaKopteva}. Using this solution we build the model of Schwarzschild-like black hole on the dust cosmological background. Solving the geodesic equations we study the motion of the test particle in the model.

The paper is organized as follows. First we briefly present the idea of the mass function method and our exact solution. Then we write the geodesic equations and solve them obtaining radial and orbital observable velocities. In conclusions section we summarize the results.

\section{The solution}
\label{sec:1}
The mass function method is the method for solving the Einstein equations by introducing the mass function \cite{MisnerSharp}-\cite{Zannias}
\begin{equation}\label{mf}
m(R,t)=r(R,t)(1+e^{-\nu(R,t)}\dot{r}^2-e^{-\lambda(R,t)}r'^2)
\end{equation}
which is one of four algebraic invariants existing for the spherically symmetric metric of general form
\begin{equation}\label{metric1}
ds^2=e^{\nu(R,t)}dt^2-e^{\lambda(R,t)}dR^2-r^2(R,t)d\sigma^2,
\end{equation}
where $d\sigma^2=d\theta^2+\sin^2\theta d\varphi^2$ is metric on unit 2-sphere. 
Using (\ref{mf}) it is possible to rewrite the Einstein equations in much simpler way
\begin{equation}
m'=\varepsilon r^2 r'; \label{ES1} 
\end{equation}
\begin{equation}
\dot{m}=-p_{\parallel} r^2 \dot{r}; \label{ES2}  
\end{equation}
\begin{equation}
2\dot{r}'=\nu ' \dot{r}+\dot{\lambda} r';\label{ES3}    
\end{equation}
\begin{equation}
2\dot{m}'=m'\frac{\dot{r}}{r'}\nu'+\dot{m}\frac{r'}{\dot{r}}\dot{\lambda}-4 r \dot{r} r' p_{\perp} .  \label{ES4}
\end{equation}
Here and further we use the system of units were velocity of light $c=1$ and factor $8 \pi G=1$; dot and prime mean partial derivatives with respect to $t$ and $R$, respectively; 
$\varepsilon$ is energy density, $p_{\perp}$ is tangential pressure, $p_{\parallel}$ is radial pressure. 

In our consideration we will use the comoving coordinates, which coincide with synchronous coordinates for the dust matter. 

The metric describing dust distribution in LTB-models \cite{Tolman} in Bonnor notations reads
\begin{equation} \label{GrindEQ__29_} 
ds^{2} =dt^{2} -\frac{r'^{2} (R,t)}{f^{2} (R)} dR^{2} -r^{2} (R,t)d\sigma ^{2} , 
\end{equation}
were $f(R)$ is arbitrary function having the sense of total energy in the shell labeled $R$ in units of $mc^2$.
The mass function in this case will take the form
\begin{equation} \label{GrindEQ__28_} 
m(R)=r(R,t)\left(1+\dot{r}^{2} (R,t)-f^{2} (R)\right). 
\end{equation} 
The dust matter is characterized by zero pressure, therefore from the equation (\ref{ES2}) it follows that the mass function depends on coordinate $R$ only.

The equation (\ref{GrindEQ__28_}) immediately leads to the Tolman solution. Indeed, expressing  $\dot{r}$ from (\ref{GrindEQ__28_}) and integrating in standard way one obtains three types of the Tolman solution depending on the sign of $f^{2} (R)-1$. We shall consider here the parabolic type of solution that implies $f^{2} (R)=1$.  
Integration gives in this case
\begin{equation} \label{GrindEQ__32_} 
r(R,t)=\left[\pm \frac{3}{2} \sqrt{m(R)} (t-t_{0} (R))\right]^{\frac{2}{3} } , 
\end{equation}
where $m(R)$ and $t_{0}(R)$ are arbitrary functions. $t_{0}(R)$ indicates the time of either initial or final singularity for each shell $R=const$. The mass function for the dust distribution has a meaning of total mass of the dust enclosed in the shell $R=const$ including gravitational energy:
\begin{equation} 
m(R)=\int_{0}^{R}\varepsilon r^{2} r' dR.
\end{equation}

The flat Friedman solution is a particular case of the solution (\ref{GrindEQ__28_}) with arbitrary functions chosen as $m(R)=a_{0}R^3$, $t_{0} (R)=0$, where $a_{0}$ is a constant concerning the present size of the universe in Friedmann models. The parabolic type of Schwarzschild solution in comoving coordinates follows from (\ref{GrindEQ__28_}) when $m(R)=r_{g}$. If then one chose $t_{0} (R)=R$ there will be known Lemaitre solution. 

Due to its physical meaning the mass function posses in some cases the property of additivity. Hence one can construct the solution for the system that may contain several sources. The solution describing the Schwarzschild-like black hole in the dust cosmological medium for the flat space case in approximation of weak interaction will be the LTB solution with combined mass function
\begin{equation} \label{GrindEQ__57_} 
m(R)=r_{g} + a_{0}R^3. 
\end{equation}

Inputting (\ref{GrindEQ__57_}) into (\ref{GrindEQ__32_}) we shall find the required solution (for the case of expansion) in the form

\begin{equation} \label{GrindEQ__58_} 
r(R,t)=\left[\frac{3}{2} \sqrt{r_{g} +a_{0} R^{3} } (t-t_{0} (R))\right]^{\frac{2}{3}} . 
\end{equation}

\begin{figure}
\begin{center}
\includegraphics{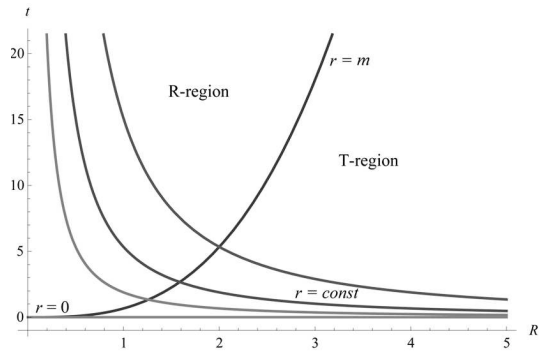}
\caption{R-T-regions for Friedmann parabolic solution. Universe starts simultaneously at every shell $R=const$. The rate of expansion in the center is slower than at distant shells. Shall-crossing singularity is absent in this solution.} 
\label{fig1}
\end{center}
\end{figure}

\begin{figure}
\begin{center}
\includegraphics{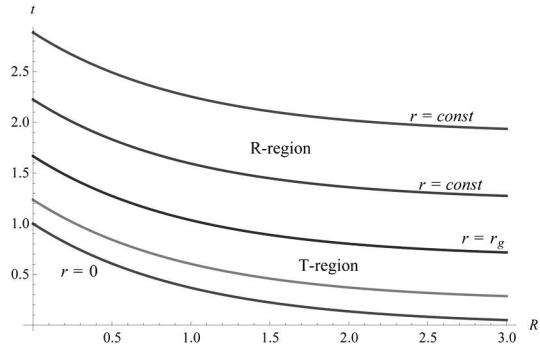}
\caption{R-T-regions for Schwarzschild parabolic solution with $t_{0}=e^{-R}$ in dimensionless units. Horizon is parallel to lines of constant 'radius'. The rate of expansion is the same for all shells. Shall-crossing singularity is absent in this solution.} 
\label{fig2}
\end{center}
\end{figure}

\begin{figure}
\begin{center}
\includegraphics{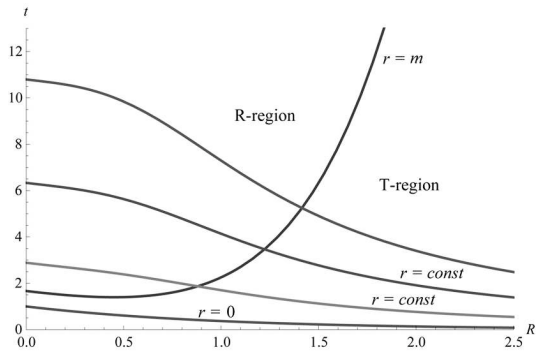}
\caption{R-T-regions for LTB parabolic solution with $t_{0}=e^{-R}$ and combined mass function in dimensionless units. In central region the behavior of horizon is similar to Schwarzschild case. In distant regions there is Fridmannian picture. Shall-crossing is absent in this case.} 
\label{fig3}
\end{center}
\end{figure}

The metric (\ref{GrindEQ__29_}) has two true singularities $r(R,t)=0$ and $r'(R,t)=0$. The first one is the initial singularity which happens when $t=t_{0}(R)$. And the second one is so called shell-crossing singularity which may happen when one shell overtakes another during the expansion. For LTB-solutions there also exists the coordinate singularity which indicates the event horizon or equivalently the boundary between R- and T-regions of the solution. It appears as a removable singularity of the metric when one switches to the curvature coordinates. But anyway it is reflected in the fact that static observer is impossible in T-region. This boundary is defined by the horizon equation
\begin{equation} \label{horizon} 
m(R)=r(R,t). 
\end{equation}
In order to avoid the shell-crossing and to satisfy the cosmic censorship conjecture we shall require
\begin{equation} \label{requirement} 
t_{sc}\leq t_{0}(R)\leq t_{hor}, 
\end{equation}
where $t_{sc}$ is a solution of the equation $r'(R,t)=0$, $t_{hor}$ is a solution of the horizon equation (\ref{horizon}). Solving these equations for the mass function (\ref{GrindEQ__57_}) and $r(R,t)$ given by (\ref{GrindEQ__58_}) one can easily get the following condition for the arbitrary function $t_{0}(R)$:
\begin{equation} \label{condition} 
\frac{2}{3}\frac{r_{g} + a_{0}R^3}{a_{0}R^2} t'_{0}(R)\leq 0. 
\end{equation}    
Constants $r_{g}$ and $a_{0}$ are positive according to their sense, and range of coordinate $R$ is $0 \leq R$ in spherical coordinate frame ($R,\theta,\varphi$), thus from (\ref{condition}) we obtain the final requirement for $t_{0}(R)$: 
\begin{equation} \label{condition2} 
t'_{0}(R)\leq 0, 
\end{equation}
that means that $t_{0}(R)$ is nonincreasing function. If we take in (\ref{GrindEQ__58_}) $t_{0}(R)=const$ we shall obtain the Friedmann solution with rescaled coordinate $R$. Consequently for our purpose to describe the inhomogeneous dust universe with embedded black hole we must require $t_{0}(R)$ to be strictly decreasing function. This physically means that for the comoving observer the universe starts at the infinity earlier then in the symmetry center $R=0$. The rate of expansion in the central region will be slower than in the periphery, similarly to Friedmann model.
Let us suppose the simplest dependence $t_{0}(R)=b e^{-R}$, where $b$ is dimensionfull constant. This is bounded monotonically decreasing function, which will not cause problems with metric coefficients at any $R$: $0 \leq R <\infty$.
To illustrate the structure of the space-time we present 2-dimensional sections in ($R,t$)-plane for Friedman solution (fig.\ref{fig1}), Schwarzschild solution (fig.\ref{fig2}) and LTB solution with combined mass function (fig.\ref{fig3}).

The energy density in the resulting space-time can be found from the equation (\ref{ES1}) regarding the mass function (\ref{GrindEQ__57_}). It reads
\begin{equation}
 \varepsilon(R,t)=\frac{4 a_{0} R^2}{\left(t-b e^{-R}\right) \left(2 b e^{-R} \left(r_{g}+a_{0} R^3\right)+3 a_{0} R^2 \left(t-b e^{-R}\right)\right)}.
 \end{equation} 

The expansion starts for each shell $R$ at the moment $t=b e^{-R}$ with infinite energy density and then for each shell  the energy density tends to zero with time. The black hole horizon appears in central region $R<<1$, but for distant shells the dynamics of the cosmological medium prevails and the picture becomes similar to Friedmann solution.

The solution (\ref{GrindEQ__58_}) with $t_{0}(R)=b e^{-R}$ can be presented in central part ($R<<1$) in the following form:
\begin{equation} \label{incenter} 
r(R<<1,t)=\left(\frac{3}{2}\right)^{2/3} r_g^{1/3} (t-b)^{2/3}+\frac{\left(\frac{2}{3}\right)^{1/3} r_g^{1/3} b R }{(t-b)^{1/3}}+O(R)^2.
\end{equation}  
The first term in this expression is exactly Schwarzschild solution (parabolic type) in comoving coordinates. Starting from the second term all the other terms  will be suppressed with time. Thus there will always be the Schwarzschild black hole in the center, which will always be hidden from the distant observer by the event horizon $r=m(R<<1)\approx r_g$. 

One can also notice here that the shell-crossing singularity is absent in the center, as far as $r'$ tends to zero at infinite time.

\section{Equations of motion}
\label{sec:1}
 In this section we shall derive the equations of motion of the test particle in obtained model, i.e. in the case when observer comoves the cosmological expansion. We shall consider here the motion in equatorial plane, so we fix  $\theta$ coordinate $\theta=\pi/2$, and chose the arbitrary function $t_{0}(R)$ as in previous section.

For the solution (\ref{GrindEQ__58_}) with mass function (\ref{GrindEQ__57_}) and metric 
\begin{equation} \label{flatmetric} 
ds^{2} =dt^{2} - r'^{2} (R,t) dR^{2} -r^{2} (R,t)d\varphi ^{2}  
\end{equation}
the geodesic equations read

\begin{equation}\label{g1}
\frac{d^2t}{ds^2}+\dot{r}'r'\Big(\frac{dR}{ds}\Big)^2+\dot{r}r \Big(\frac{d\varphi}{ds}\Big)^2 = 0; 
\end{equation}

\begin{equation}\label{g2}
\frac{d^2R}{ds^2}+ \frac{r''}{r'}\Big(\frac{dR}{ds}\Big)^2-\frac{r}{r'}\Big(\frac{d\varphi}{ds}\Big)^2+2\frac{\dot{r}'}{r'}\frac{dt}{ds}\frac{dR}{ds}=0; 
\end{equation}

\begin{equation}\label{g3}
\frac{d^2\varphi}{ds^2}+2\frac{r'}{r}\frac{dR}{ds}\frac{d\varphi}{ds}+2\frac{\dot{r}}{r}\frac{d\varphi}{ds}\frac{dt}{ds}=0. 
\end{equation}

For the case when the particle starts from rest with respect to comoving coordinates $R$, $\varphi$ one has 
\begin{equation}
\frac{dR}{ds} = 0, \quad \frac{d\varphi}{ds}=0, \quad \frac{dt}{ds}=1,
\end{equation}
and hence from the system (\ref{g1})-(\ref{g3}) it follows that
\begin{equation}
\frac{d^2t}{ds^2}=0,\quad \frac{d^2R}{ds^2}=0, \quad \frac{d^2\varphi}{ds^2}=0.
\end{equation}

This means that starting from rest the particle is staying in rest and follows the cosmological expansion as all matter averagely do.

To investigate the more general case when the particle has arbitrary initial velocity in $\theta=\pi/2$ plane it is more convenient to use the equations (\ref{g2}) and (\ref{g3}) taking into account the interval (\ref{flatmetric}). Let us first introduce the following notations
\begin{equation}\label{notations1}
 \frac{dR}{dt}\equiv v, \quad \frac{d\varphi}{dt}\equiv \omega, \quad \frac{dt}{ds}\equiv x;
\end{equation}
\begin{equation}\label{notations2}
 u_1\equiv r'v, \quad   u_3\equiv r\omega,
\end{equation}
where $u_1$ and $u_3$ are observable radial and orbital velocities of the test particle, respectively.

In terms of notations (\ref{notations1}) the equation (\ref{g2}) may be rewritten in the following form
\begin{equation}\label{g22}
 \frac{1}{x}\frac{dx}{dt}=-\frac{1}{v}\frac{dv}{dt}-\frac{1}{r'}\left(\frac{\partial r'}{\partial R}v + \frac{\partial r'}{\partial t}\right)-\frac{1}{r'}\frac{\partial r'}{\partial t} + \frac{r}{r'}\frac{\omega^2}{v}.
\end{equation}
The integration of (\ref{g22}) gives
\begin{equation}\label{int1}
\frac{ds}{dt}=C(R)r'u_1e^{-\int\frac{u_3}{u_1}d\varphi},  
\end{equation}
where $C(R)$ is arbitrary function of integration.

The similar procedure applied to the equation (\ref{g3}) will give
\begin{equation}\label{g32}
 \frac{1}{x}\frac{dx}{dt}=-\frac{1}{\omega}\frac{d\omega}{dt}-\frac{2}{r}\left(\frac{\partial r}{\partial R}v + \frac{\partial r}{\partial t}\right),
\end{equation} 
and after integration there will be
\begin{equation}
\frac{ds}{dt}= Aru_3, \label{int2}
\end{equation}
where $A$ is arbitrary constant of integration.

Combining together (\ref{int1}) and (\ref{int2}) and taking the logarithm one will obtain
\begin{equation}\label{combint}
-\int\frac{u_3}{u_1}d\varphi=\ln{\frac{Ar}{C(R)r'}}+\ln{\frac{u_3}{u_1}}.
\end{equation} 
Taking the derivative with respect to $\varphi$ from both sides of (\ref{combint}) one will get the equation that leads to the simple result
\begin{equation}\label{result}
\frac{u_3}{u_1}=\varphi+B,
\end{equation}
with $B$ being arbitrary constant of integration.

From the expression for the interval (\ref{flatmetric}) it follows that
\begin{equation}
\Big(\frac{ds}{dt}\Big)^2=1-(u_1^2+u_3^2). \label{int3}
\end{equation}

Combining (\ref{int2}) and (\ref{int3}) one finds for the orbital velocity
\begin{equation} \label{u3}
u_3=\pm\frac{\sqrt{1-u_1^2}}{\sqrt{1+A^2r^2}}.
\end{equation} 

And hence the radial velocity will read
\begin{equation} \label{u1}
u_1=\pm\frac{\varphi+B}{\sqrt{1+A^2r^2+\left(\varphi+B\right)^2}}.
\end{equation}   
The total observable velocity of the test particle can be found in standard way $u^2=u_1^2+u_3^2$. This is the velocity which would be measured with usual instruments by the observer being in rest with respect to cosmological medium. 

Let us suppose that at some time $t_{in}$ when we start to observe the particle, it occupies coordinates $R_{0}$ and $\varphi_{0}$ with respective values of $r(R_{0},t_{in})\equiv r_{0}$ and $r'(R_{0},t_{in})\equiv r'_{0}$. And let the initial velocity of the particle with its components will be $u_{0}^2=u_{01}^2+u_{03}^2$. The important condition here is  $t_{in}>be^{-R_{0}}$, that means that by the moment $t_{in}$ the universe has already started in the shell $R=R_{0}$. 

Using these initial conditions we find from the equations (\ref{u3}), (\ref{u1}) the values of $A$ and $B$. They are   
\begin{equation}
A=\frac{\sqrt{1-u_{0}^2}}{r_{0}u_{03}}, \quad B=\frac{u_{01}}{u_{03}}-\varphi_{0}.
\end{equation}

And finally for the velocities onehas
\begin{equation}\label{u12}
u_1=\pm\frac{r_0 (u_{01}+u_{03}(\varphi -\varphi_0))}{\sqrt{r^2 (1-u_{0}^2)+r_{0}^2 \left(u_{03}^2+(u_{01}+u_{03}(\varphi -\varphi_0))^2 \right)}}
\end{equation}
\begin{equation}\label{u32}
u_3=\pm\frac{r_0 u_{03}}{\sqrt{r^2 (1-u_{0}^2)+r_{0}^2 \left(u_{03}^2+(u_{01}+u_{03}(\varphi -\varphi_0))^2 \right)}}
\end{equation}

\begin{figure}
\begin{center}
\includegraphics{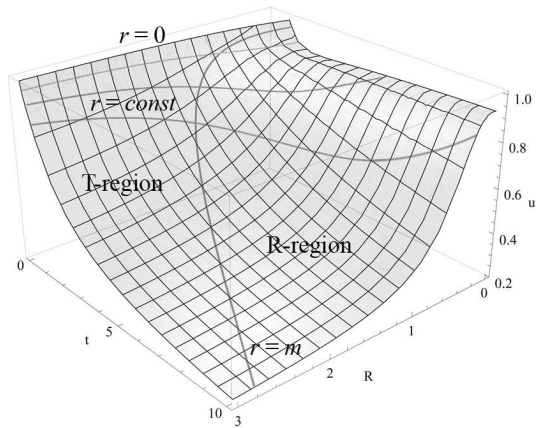}
\caption{Total velocity profile of the test particle. $A=B=1$, $\varphi=1$. $R$, $t$ and $u$ are given in dimensionless units. $u=1$ is speed of light.} 
\label{profile}
\end{center}
\end{figure}

To illustrate the picture seen by the observer we present the surface profile of the total velocity at fig.\ref{profile}.  

From the point of view of such observer the particle will lose its velocity with time and will be involved to the cosmological expansion. At R-region it is possible to introduce the static distant observer who probably could see the black hole horizon in central region and could investigate the possibility for stationary orbits. Our chosen comoving coordinate frame does not allow to study such questions. The energy and momentum are not the integrals of motion here. All we can see is how particles move out of the singularity or how they fall down (in case we chose another sign of the solution).  

\section{Conclusions}
\label{sec:1}
In this work we constructed the model of the black hole enclosed into the dust cosmological background in case of zero spatial curvature. This model is based on our exact solution (\ref{GrindEQ__58_}) of the class of LTB inhomogeneous solutions. We described the properties of the model and built the R-T-structure of the resulting space-time. It was shown that central region includes the Schwarzchild-like part of horizon (see fig.\ref{fig3}) and there always be a black hole in the center (\ref{incenter}). We derived the equations of motion (\ref{g1})-(\ref{g3}) of the test particle from the point of view of the observer comoving to cosmological expansion. We found analytical expressions (\ref{u32}), (\ref{u12}) for observable orbital and radial velocities of the particle and plotted the surface profile of the total velocity in this case (see fig.\ref{profile}). In comoving coordinate frame it is impossible to study the questions concerning the black hole horizon but one can observe the local motion of the particles influenced by the cosmological expansion.

\end{document}